\documentclass[prl,aps,float,nofootinbib,superscriptaddress,twocolumn]{revtex4-1}
\usepackage{amssymb}
\usepackage{amsmath}
\usepackage{amsfonts}
\usepackage{epsfig}
\usepackage{bbm}
\usepackage{comment}
\usepackage{multirow}
\usepackage{longtable}
\usepackage{array}
\usepackage{subfigure}
\usepackage{graphics}
\usepackage{array}
\usepackage[usenames,dvipsnames,svgnames,table]{xcolor}
\usepackage{amsxtra}
\usepackage{dsfont} 
\usepackage{enumitem}

\definecolor{bluebb}{RGB}{46,84,206}
\definecolor{redbb}{RGB}{228,30,43}
\definecolor{greenbb}{RGB}{34,139,34}

\newcommand{\bra}[1]{\langle #1 \vert}
\newcommand{\ket}[1]{\vert #1 \rangle}
\newcommand{\scal}[2]{\langle #1 \vert #2 \rangle}
\newcommand{\elma}[3]{\bra{#1} #2 \ket{#3}}

\DeclareMathOperator{\Tr}{Tr}

\allowdisplaybreaks
\begin{document}
\title{Calculating the norm matrix to solve the A-body Schr\"{o}dinger equation within a set of non-orthogonal many-body states}

\author{B. Bally}
\affiliation{ESNT, IRFU, CEA, Universit\'e Paris - Saclay, F-91191 Gif-sur-Yvette, France}

\author{T. Duguet}
\email{thomas.duguet@cea.fr} 
\affiliation{IRFU, CEA, Universit\'e Paris - Saclay, F-91191 Gif-sur-Yvette, France}
\affiliation{KU Leuven, Instituut voor Kern- en Stralingsfysica, 3001 Leuven, Belgium}
\affiliation{National Superconducting Cyclotron Laboratory and Department of Physics and Astronomy, Michigan State University, East Lansing, MI 48824, USA}

\date{\today}

\begin{abstract}
There are efficient many-body methods, such as the (symmetry-restored) generator coordinate method in nuclear physics, that formulate the A-body Schr\"{o}dinger equation within a set of non-orthogonal many-body states. Solving the corresponding secular equation requires the evaluation of the norm matrix and thus the capacity to compute its entries consistently and without any phase ambiguity. This is not always a trivial task, e.g.\@ it remained a long-standing problem for methods based on general Bogoliubov product states. While a solution to this problem was found recently in Ref.~[L. M. Robledo, Phys. Rev. C79, 021302 (2009)], the present work introduces an alternative method that can be generically applied to other classes of states of interest in many-body physics. The method is presently exemplified in the case of Bogoliubov states and numerically illustrated on the basis of a toy model.
\end{abstract}
\maketitle

%
%

{\it Introduction}. The A-body Schr\"{o}dinger equation is most often (approximately) solved by representing it on a (truncated) orthonormal basis of the A-body Hilbert space $\mathcal{H}_A$. Some approaches, however, represent the A-body Schr\"{o}dinger equation on a finite set of  {\it non-orthogonal} states of ${\cal H}_A$. At the price of losing the orthogonality, it permits to select fewer states on the basis of their expected physical relevance. The set may even exceed ${\cal H}_A$ by employing states mixing vectors belonging to Hilbert spaces associated with different particle numbers, i.e.\@ states that are genuine vectors of Fock space ${\cal F}$. This is for instance the case of the generator coordinate method (GCM) in use in nuclear physics~\cite{ring80a,bender03b,Duguet:2013dga,Egido:2016bdz} and of symmetry  restoration calculations employed in nuclear physics~\cite{ring80a,bender03b,Duguet:2013dga,Egido:2016bdz} and quantum chemistry~\cite{jimenez12a,jimenez12b,samanta12a,rivero13a}. In both methods, the Hamilton operator is diagonalized in the finite-dimensional vector space spanned by a set of non-orthogonal Bogoliubov product states. The secular equation to be solved 
requires the evaluation of the norm matrix constructed from overlaps between all members of the set. Furthermore, the recently developped particle-number-restored Bogoliubov 
coupled-cluster and particle-number-restored many-body perturbation theories~\cite{Duguet:2015yle} also build on a manifold of non-orthogonal Bogoliubov states. The norm kernels at play are more general as they explicitly incorporate many-body correlations and reduce to the mere overlap between two non-orthogonal Bogoliubov states whenever such correlations 
are omitted. Last but not least, the efficient computation of overlaps constitutes a key element of quantum monte carlo (QMC) approaches, especially when they rely on more elaborate walkers and/or trial states than Slater determinants~\cite{guerrero98a,roggero13a,puddu03a,juillet17a}.

The closed-form evaluation of the overlap between many-body states constitutes a long-standing problem. For instance, while the overlap between two non-orthogonal A-body Slater determinants has long been known to be computable as a determinant~\cite{blaizot86}, the efficient and unambiguous\footnote{In general, the well-known Onishi formula~\cite{onishi66} can provide the norm of the overlap but not its complex phase.} calculation of the overlap between two arbitrary Bogoliubov quasiparticle states has only become available recently as the Pfaffian of a skew-symmetric matrix~\cite{Robledo:2009yd,Robledo:2011ce,Avez:2011wr}. 

In this context, we presently propose a closed-form expression for the overlap between arbitrary many-body states and implement it in the context of GCM calculations based on a set of Bogoliubov states. All pertinent technical details related to the latter case can be found in Ref.~\cite{Bally:2017nom}. The application to other classes of states in use in many-body physics remains to be investigated in the future. 

{\it Objective}. We consider the situation where the static A-body Schr\"{o}dinger equation is represented on a set of $N$ (a priori non-orthogonal) many-body states
\begin{equation}
\mathcal{M} \equiv \big\{ \ket{\Phi_{k}}, k = 1\ldots, N\big\} \, . \label{manifold}
\end{equation}
The eigenvectors of the Hamiltonian $H$ are expanded in terms of the states of $\mathcal{M}$ according to
\begin{equation}
 \ket{\Psi_n} \equiv \sum_{k=1}^N f_{nk} \ket{\Phi_{k}} \, , 
\end{equation}
where $f_{nk}$ are complex numbers. The corresponding secular equation takes the form of a generalized eigenvalue problem
\begin{equation}
{\cal H} \mathfrak{f}_n = E_n \, {\cal N} \mathfrak{f}_n \, , \label{secular}
\end{equation}
where $\mathfrak{f}_n$ is the weight column matrix, i.e.\@ $(\mathfrak{f}_n)_k \equiv f_{nk}$, and where the norm and Hamiltonian hermitian matrices are made out of $N(N+1)/2$ independent elements 
\begin{subequations}
\label{matrixelements}
\begin{align}
{\cal N}_{kl} &\equiv \langle \Phi_{k} | \Phi_{l} \rangle \, , \label{matrixelements1} \\
{\cal H}_{kl} &\equiv \langle \Phi_{k} | H | \Phi_{l} \rangle \, , \label{matrixelements2}
\end{align}
\end{subequations}
respectively. Solving Eq.~\eqref{secular} gives access to energies $E_n$ and weigths $\mathfrak{f}_n$ for states $\ket{\Psi_n}$. Typically, such a generalized eigenvalue
problem is addressed by first diagonalizing the norm matrix ${\cal N}$ before solving a standard eigenvalue problem in the (possibly smaller) orthonormal basis built from the norm eigenvectors
with non-zero eigenvalues. A key step of the many-body calculation is thus the consistent computation of the $N(N+1)/2$ independent entries of the norm matrix, including their complex phase.

{\it Master formulae}. Given an arbitrary pair of states $(\ket{\Phi_{k}},\ket{\Phi_{l}})$ belonging to $\mathcal{M}$, we assume that a unitary transformation linking both states is either known or can be extracted under the form
\begin{equation}
| \Phi_{l} \rangle   \equiv e^{iS[k,l]} |  \Phi_{k} \rangle   \, , \label{unitary transfor}
\end{equation}
where $S[k,l]$ is a hermitian operator.
Based on this sole hypothesis, an auxiliary set connecting both states is introduced as
\begin{equation}
\mathcal{M}[k,l] \equiv \big\{ \ket{\Phi_{kl}(\theta)} \equiv e^{i \theta S[k,l]}  |  \Phi_{k} \rangle \, , \, \theta \in [0, 1] \big\} \, ,
\end{equation}
such that $\ket{\Phi_{kl}(0)}=\ket{\Phi_k}$  and $\ket{\Phi_{kl}(1)} = | \Phi_l \rangle$. Considering an {\it arbitrary} many-body bra $\langle \Theta |$, the quantity
\begin{equation}
n_{kl}[\langle \Theta |, \theta] \equiv \frac{\langle \Theta | \Phi_{kl}(\theta) \rangle }{\langle \Theta | \Phi_k \rangle} 
\end{equation}
satisfying $n_{kl}[\langle \Theta |, 0]=1$ is defined along the auxiliary manifold. Differentiating it with respect to $\theta$ leads to
\begin{equation}
\label{intermediate1}
\begin{split}
 \frac{d}{d\theta} \, n_{kl}[\langle \Theta |, \theta] &= i \frac{\elma{\Theta}{S[k,l]}{\Phi_{kl}(\theta)}}{\langle \Theta | \Phi_k \rangle} \, .
\end{split}
\end{equation}
Assuming that $n_{kl}[\langle \Theta |, \theta] \ne 0$ along $\mathcal{M}[k,l]$, one divides both sides of Eq.~\eqref{intermediate1} by it and integrates the corresponding first-order differential equation between $0$ and $\theta$ to obtain
\begin{equation}
\label{eq:genov}
n_{kl}[\langle \Theta |, \theta] = e^{i\int_0^\theta d\phi \, s_{kl}[\langle \Theta |, \phi]} \, ,
\end{equation}
where the off-diagonal {\it linked-connected} kernel~\cite{Duguet:2014jja,Duguet:2015yle} of the operator $S[k,l]$ defined along the manifold $\mathcal{M}[k,l]$
\begin{equation}
s_{kl}[\langle \Theta |, \theta] \equiv  \frac{\elma{\Theta}{S[k,l]}{\Phi_{kl}(\theta)}}{\scal{\Theta}{\Phi_{kl}(\theta)}} \, , \label{offdiagonaloperatorkernel}
\end{equation}
is unambiguous in the sense that it is independent of the relative phase between $| \Phi_k \rangle$ and $| \Phi_l \rangle$. Equations~\eqref{eq:genov}-\eqref{offdiagonaloperatorkernel} constitute the master formulae repeatedly used below to access the norm matrix (Eq.~\eqref{matrixelements1}).

{\it Phase convention}. The normalized states belonging to $\mathcal{M}$ are all individually defined up to a phase that must not influence the computation of observables. This freedom must be explicitly controlled such that the $N(N+1)/2$ independent entries to the norm matrix are computed consistently. This relates to fixing the relative phases of the states in a synchronized fashion, which effectively impacts the definition of each operator $S[k,l]$. This can be done by specifying the phase each member of $\mathcal{M}$ entertains with a common known state of reference generically denoted as $| \bar{\Phi} \rangle$. Among many possibilities, a natural and practical choice consists of picking this pivot state within $\mathcal{M}$ itself and requiring that all states of $\mathcal{M}$ have the same phase relative to it. Accordingly, we choose $\ket{\bar{\Phi}}\equiv \ket{\Phi_{1}}$, although any other state of $\mathcal{M}$ would be equally appropriate, and impose that
\begin{equation}
\text{Arg} (\langle \Phi_{1} | \Phi_{1} \rangle) = \text{Arg} (\langle \Phi_{1} | \Phi_{2} \rangle) \ldots = \text{Arg} (\langle \Phi_{1} | \Phi_{N} \rangle) = 0 \, , \nonumber
\end{equation}
given that $\langle \Phi_{1} | \Phi_{1} \rangle$ is real. 

{\it Algorithm}. The evaluation of the $N(N+1)/2$ independent entries to the norm matrix follows three successive steps
\begin{enumerate}
\item The $N$ diagonal elements ${\cal N}_{kk}$ are trivially obtained by normalizing all members of the set, i.e.\@ by imposing that $\langle \Phi_{k} | \Phi_{k} \rangle=1$ for $k=1, \ldots, N$. 
\item The $N-1$ remaining elements of the first row are computed on the basis of the  $N-1$ operators $S[1,l]$.  Applying Eq.~\eqref{eq:genov} for $k=1$, $\langle \Theta | \equiv \langle \Phi_1 |$ and $\theta=1$, one obtains
\begin{equation}
\label{eq:genovHFBNO1l_A}
\frac{{\cal N}_{1l}}{{\cal N}_{11}} = e^{i\int_0^1 d\phi \, s_{1l}[\langle \Phi_1 |, \phi]}  \, ,
\end{equation}
where $s_{1l}[\langle \Phi_1 |, \theta]$ runs over the manifold $\mathcal{M}[1,l]$ and is defined for the bra $\langle \Phi_1 |$. The phase convention stated above constrains the pure number entering the definition of $S[1,l]$ according to
\begin{equation}
\Re e  \int_0^1 d\phi \, s_{1l}[\langle \Phi_1 |, \phi] = 0 \, , \label{phaseconvention}
\end{equation}
such that ${\cal N}_{1l}$ is effectively real. 
\item Applying Eq.~\eqref{eq:genov} again for $k=1$ but now setting $\langle \Theta | \equiv \langle \Phi_m |$, $1<m<l\leq N$, the remaining $(N-1)(N-2)/2$ independent overlaps ${\cal N}_{ml}$ are obtained consistently for $\theta=1$ via
\begin{equation}
\label{eq:genovHFBNOml}
\frac{{\cal N}_{ml}}{{\cal N}_{m1}} =  e^{i\int_0^1 d\phi \, s_{1l}[\langle \Phi_m |, \phi]} \, ,
\end{equation}
where the off-diagonal kernel of $S[1,l]$ still runs over the manifold $\mathcal{M}[1,l]$ but now involves the bra $\langle \Phi_m |$ rather than $\langle \Phi_1 |$. Since ${\cal N}_{m1}={\cal N}_{1m}$ is among the $N-1$ overlaps computed in step 2, Eq.~\eqref{eq:genovHFBNOml} completes the norm matrix. While the diagonal, the first row and the first column of the norm matrix are real, the remaining entries are a priori complex.
\end{enumerate}

Each entry ${\cal N}_{ml}$, including its complex phase, has been powerfully expressed in terms of the integral of an off-diagonal kernel of the operator $S[1,l]$ along the auxiliary manifold connecting $| \Phi_l \rangle$ to $| \Phi_1 \rangle$. In essence, the rationale of the method is to commute the computation of a pure overlap into the computation of the linked-connected kernel of an operator that has no phase ambiguity. Of course, the usefulness of the method relies on our ability to compute such an operator kernel, which itself depends on the nature of the many-body states making up $\mathcal{M}$. More specifically, the character of $| \Phi_{1}\rangle$ and $| \Phi_{l}\rangle$ determines the {\it nature} of the operator $S[1,l]$ driving their unitary connection along with our capacity to extract it and compute efficiently its off-diagonal kernels $s_{1l}[\langle \Phi_m |, \theta]$.

{\it Application}. We apply the above scheme on the basis of a set made out of arbitrary Bogoliubov states~\cite{ring80a,blaizot86}. This particular choice  is characteristic of state-of-the-art (symmetry-restored) GCM calculations in nuclear physics. 

Each member $| \Phi_{k}\rangle$ of $\mathcal{M}$ is defined as a vacuum, i.e.\@ $\forall \, \mu, \beta^{[k]}_{\mu} | \Phi_{k}\rangle = 0$, of the set of quasi-particle operators  $\{ \beta^{[k]}_{\mu} ; \beta^{[k]\dagger}_{\mu} \}$. These  creation and annihilation operators relate to those defining a basis $\{ c_p ; c^{\dagger}_p \}$ of the $n$-dimensional one-body hilbert space ${\cal H}_1$ via a linear Bogoliubov transformation
\begin{align}
\begin{pmatrix}
\beta \\
\beta^{\dagger}
\end{pmatrix}_{[k]}
&\equiv {\cal W}^{\dagger}_{[k]}
\begin{pmatrix}
c \\
c^{\dagger}
\end{pmatrix}   
\equiv 
\begin{pmatrix}
U^{\dagger} & V^{\dagger} \\
V^{T} &  U^{T}
\end{pmatrix}_{[k]}
\begin{pmatrix}
c \\
c^{\dagger}
\end{pmatrix}   
\, ,
\end{align}
where the unitarity of the $2n\times 2n$ matrix ${\cal W}_{[k]}$ ensures the fermionic character of the quasi-particle operators. The above procedure actually defines $| \Phi_{k}\rangle$ only up to a complex phase~\cite{hara79a,blaizot86}.

Given $| \Phi_{k}\rangle$ and $| \Phi_{l}\rangle$, the operator $S[k,l]$ parameterizing their unitary connection is a generic hermitian {\it one-body} operator reading, in the quasi-particle basis of an arbitrary third state $| \Phi_{m}\rangle$, as
\begin{align}
S[k,l] &\equiv S^{00}[k,l]_{[m]} + \frac12 \Tr\left(S^{11}[k,l]_{[m]}\right) \label{eq:1bodynobogok} \\
&+ \frac12  
\left(\,\beta^{\dagger} \hspace{0.1cm} \beta \,\right)_{[m]}
\begin{pmatrix}
S^{11}[k,l] & S^{20}[k,l] \\
-S^{02}[k,l] &  -S^{11 \ast}[k,l]
\end{pmatrix}_{[m]}
\begin{pmatrix}
\beta \\
\beta^{\dagger}
\end{pmatrix} _{[m]} \nonumber
\end{align}
where $S^{00}[k,l]_{[m]}$ is a real number, $S^{11}[k,l]_{[m]}$ is a hermitian matrix whereas $S^{20}[k,l]_{[m]}$ and $S^{02}[k,l]_{[m]}$ denote skew-symmetric matrices satisfying \mbox{$S^{02}[k,l]_{[m]}=(S^{20}[k,l]_{[m]})^{\ast}$}. The non-trivial part of the operator $S[k,l]$ is uniquely extracted~\citep{Bally:2017nom} from the sole knowledge of the Bogoliubov transformations ${\cal W}_{[k]}$ and ${\cal W}_{[l]}$, i.e.\@ it reads in the quasi-particle basis of  $| \Phi_{k}\rangle$ as
\begin{align}
\begin{pmatrix}
S^{11}[k,l] & S^{20}[k,l] \\
-S^{02}[k,l] &  -S^{11 \ast}[k,l]
\end{pmatrix}_{[k]} 
&=  i \log({\cal W}^{\dagger}_{[l]}{\cal W}_{[k]}) \, .
\end{align}

Following the algorithm layed down above, only the $N-1$ operators $S[1,l]$ are effectively needed. Their off-diagonal kernels $s_{1l}[\langle \Phi_m |, \theta]$ along the auxiliary manifold $\mathcal{M}[1,l]$ can be unambiguously computed on the basis of the off-diagonal Wick theorem~\cite{balian69a}. Fixing the constant $S^{00}[1,l]_{[1]}$ entering $S[1,l]$ via the application of Eq.~\eqref{phaseconvention}, the entries on the first row of the norm matrix (Eq.\eqref{eq:genovHFBNO1l_A}) are thus obtained under the form
\begin{equation}
\frac{{\cal N}_{1l}}{{\cal N}_{11}} = e^{ -\Im m \frac{1}{2} \int_0^1 d\phi \Tr \left( S^{02}[1,l]_{[1]} R^{--}_{1l}[\langle \Phi_1 |, \phi] \right)}  \, ,
\end{equation}
where the elementary off-diagonal contraction defined as
\begin{equation}
\label{elementarycontract}
\left(R^{--}_{1l}[\langle \Phi_1 |, \phi] \right)_{k_1k_2} \equiv  \frac{\langle \Phi_1 | \beta^{[1]}_{k_1}\beta^{[1]}_{k_2} | \Phi_{1l}(\theta) \rangle}{\langle \Phi_1 | \Phi_{1l}(\theta) \rangle}   \, ,
\end{equation}
is computable from ${\cal W}_{[1]}$ and ${\cal W}_{[l]}$~\citep{Bally:2017nom}. Eventually, the norm matrix can be completed via a similar specification of Eq.~\eqref{eq:genovHFBNOml} that requires the introduction of the elementary contraction $R^{--}_{1l}[\langle \Phi_m |, \phi]$. The Onishi formula~\cite{onishi66} is recovered for each entry by taking the norm of the corresponding expression~\cite{Bally:2017nom}. 

The overlap between two arbitrary Bogoliubov product states can be alternatively computed as the Pfaffian of a skew-symmetric matrix~\cite{Robledo:2009yd,Robledo:2011ce,Avez:2011wr}. This result relies on the Thouless representation~\cite{thouless60} of the Bogoliubov states and, as such, relates to a phase convention, i.e.\@ $\forall \, k, \text{Arg} (\langle 0 | \Phi_k \rangle) = 1$, that differs from the one presently used. This is interesting to demonstrate that, while impacting individual overlaps, the overall phase convention does not influence output observables. All is needed is an internally consistent computation of the entries to the norm and Hamiltonian matrices. 

{\it Toy model}. The method is numerically implemented on the basis of a set $\mathcal{M} \equiv \big\{ | \Phi_1 \rangle, | \Phi_2 \rangle, | \Phi_3 \rangle\big\}$ of three different Bogoliubov states. The associated norm matrix reads as
\begin{equation}
{\cal N} \equiv 
\begin{pmatrix}
\langle \Phi_{1} | \Phi_{1} \rangle & \langle \Phi_{1} | \Phi_{2} \rangle & \langle \Phi_{1} | \Phi_{3} \rangle \\
\langle \Phi_{2} | \Phi_{1} \rangle & \langle \Phi_{2} | \Phi_{2} \rangle & \langle \Phi_{2} | \Phi_{3} \rangle \\
\langle \Phi_{3} | \Phi_{1} \rangle & \langle \Phi_{3} | \Phi_{2} \rangle & \langle \Phi_{3} | \Phi_{3} \rangle 
\end{pmatrix} 
  \label{normmatrixGCM3states} \, .
\end{equation}
The Bogoliubov transformation associated with $| \Phi_{k} \rangle$ presently reads as
\begin{align}
{\cal W}_{[k]} &\equiv {\cal L}_{[k]} \, \bar{{\cal W}}_{[k]} \equiv
\begin{pmatrix}
L & 0 \\
0 &  L^{\ast}
\end{pmatrix}_{[k]}
\begin{pmatrix}
\bar{U} & \bar{V}^{\ast} \\
\bar{V} &  \bar{U}^{\ast}
\end{pmatrix}_{[k]}  
  \,\, ,
\end{align}
where $L$ denotes a random $n\times n$ complex unitary matrix transforming the ten-dimensional, i.e.\@ $n=10$,  basis of ${\cal H}_1$ made out of five doubly-degenerated single-particle levels. Following ${\cal L}_{[k]}$, $\bar{{\cal W}}_{[k]}$ is a BCS transformation characterized by the set of real $2\times 2$ blocks of the form
\begin{subequations}
\label{BCStransforphi}
\begin{align}
\bar{U}_{[k]}(p,\bar{p}) &\equiv
\begin{pmatrix}
+u_p[k] & 0 \\
0 &  +u_p[k]
\end{pmatrix} \,\, , \\
\bar{V}_{[k]}(p,\bar{p}) &\equiv
\begin{pmatrix}
0 & +v_p[k] \\
-v_p[k] &  0
\end{pmatrix}  \,\, ,
\end{align}
\end{subequations}
where $\bar{p}$ denotes the conjugated partner of $p$ and where $u^{2}_p[k]+v^{2}_p[k]=1$. The BCS occupations $v^{2}_p[k]$, $p=1,\ldots 5$, are decreasingly chosen in the interval $]0,1[$ to mimic a realistic fully paired system. Even though, for the sake of simplicity, fully paired states are presently considered, the validity of the method has been checked for Bogoliubov states containing an even or odd number of fully occupied single-particle states.

\begin{figure}
\begin{center}
\includegraphics[width=8.0cm]{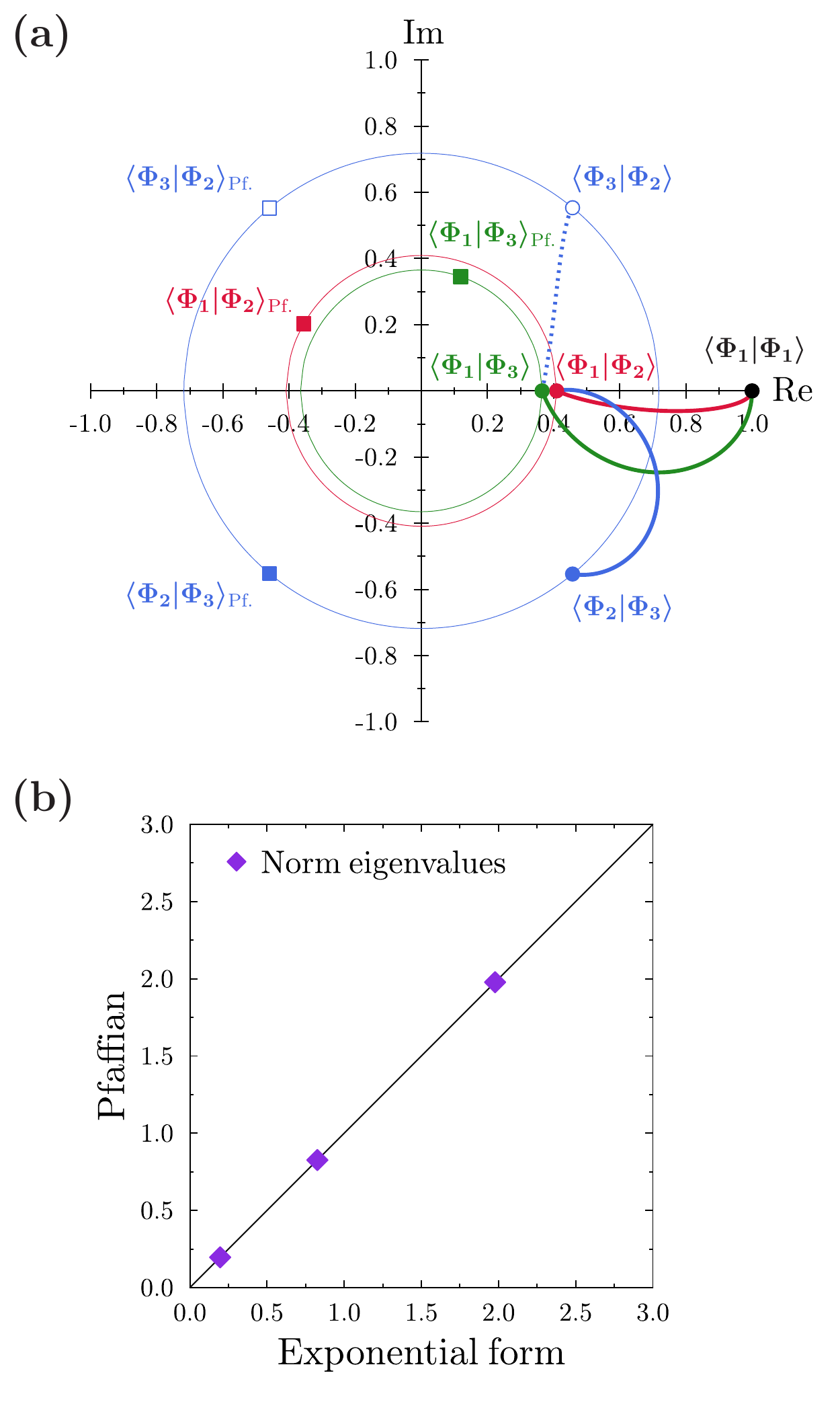}
\end{center}
\caption{(Color online) Upper panel: complex overlaps making up the $3 \times 3$ norm matrix represented. Squares denote values obtained from the Pfaffian method whereas circles denote those obtained from the present method. The lines provide the auxiliary pathes followed from one overlap to the other, starting from $\langle \Phi_1 | \Phi_1 \rangle = 1$. Lower panel: eigenvalues of the norm matrix obtained on the basis of the Pfaffian method against those obtained with the present method. }
\label{results10levelsGCMBogomodel}
\end{figure}

\begin{figure}
\begin{center}
\includegraphics[width=7.5cm]{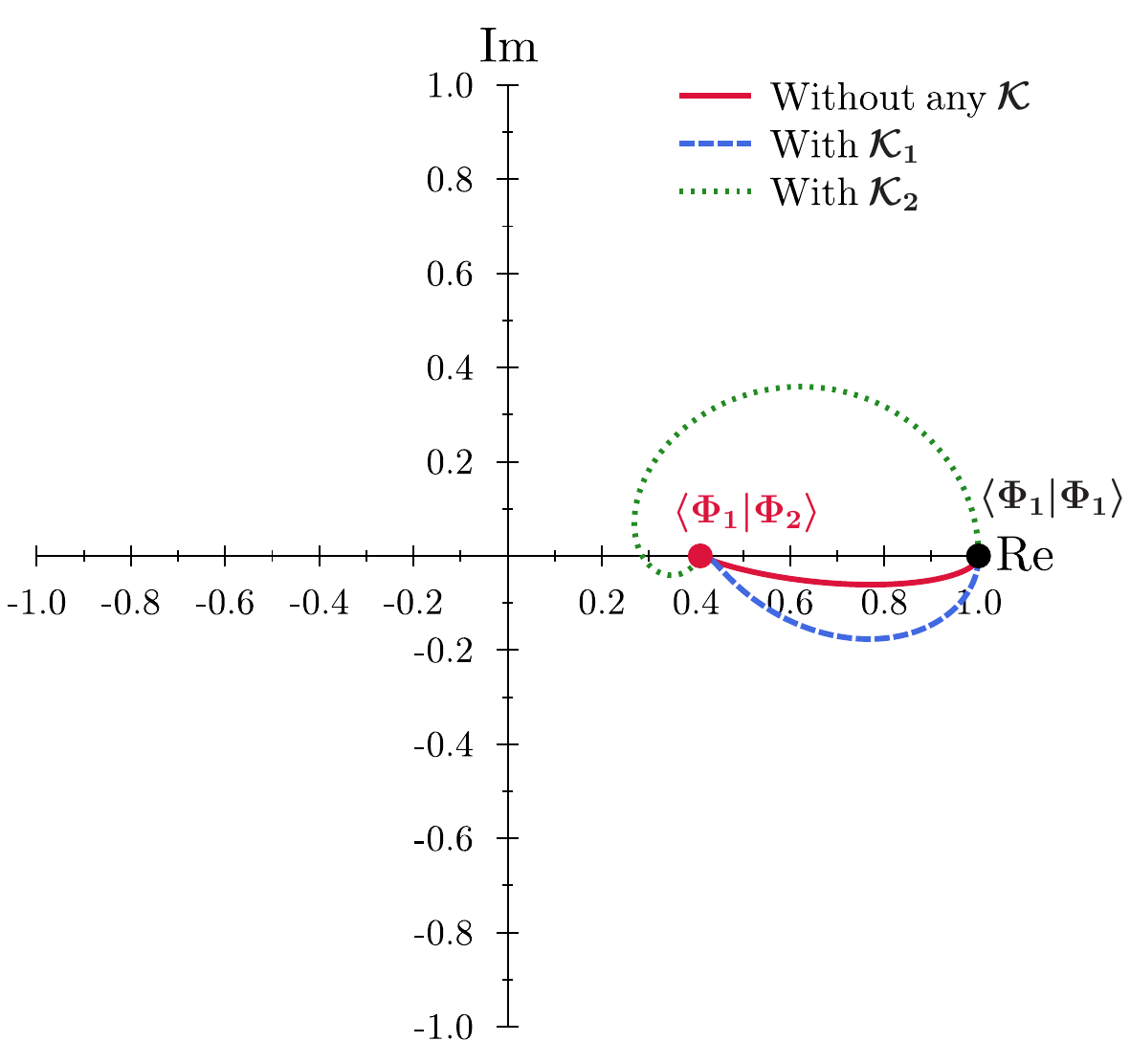}
\end{center}
\caption{(Color online) Overlap $\langle \Phi_1 | \Phi_2 \rangle$ represented in the complex plane. The color lines correspond to the increment integration along the auxiliary manifolds linking both states obtained without (full red line) and with (dashed blue and dotted green lines) additional trivial Bogoliubov transformations ${\cal K}_1$ and ${\cal K}_2$. The  $n \times n$ unitary matrices $K_1$ and  $K_2$ are randomly generated. As many different pathes as required can be generated in this way.}
\label{results10levelsBCSmodel}
\end{figure}

{\it Results}. The upper panel of Fig.~\ref{results10levelsGCMBogomodel} displays in the complex plane individual overlaps making up the norm matrix. Squares represent the results obtained from the Pfaffian method whereas circles denote those computed from the presently proposed method. Thick lines provide the auxiliary pathes followed from one overlap to the other, starting from $\langle \Phi_1 | \Phi_1 \rangle = 1$. Consistently with the scheme exposed above, all the overlaps involving the pivot state $| \Phi_1 \rangle$ are real, which is not the case for the Pfaffian method. Interestingly, the complex conjugate values $\langle \Phi_2 | \Phi_3 \rangle$ and $\langle \Phi_3 | \Phi_2 \rangle$ are consistently obtained by going through $| \Phi_2 \rangle$ or $| \Phi_3 \rangle$ first. Three circles help visualize that, while individual overlaps differ in both methods because of the distinct phase conventions used, they only do so by a complex phase. Eventually, the lower panel demonstrates that the eigenvalues of the norm matrix obtained from both methods are identical, thus showing the consistency of both calculations and the independence of the result on the phase convention used.

The derivation of Eq.~\eqref{eq:genov} relied on the hypothesis that $n_{kl}[\langle \Theta |, \theta] \ne 0$ along $\mathcal{M}[k,l]$. In the present application, it is clear from Fig.~\ref{results10levelsGCMBogomodel} that this hypothesis is indeed fulfilled for all the overlaps involved. There however exists situations, e.g.\@ global-gauge symmetry restoration calculations, in which it is not the case~\cite{Bally:2017nom}. To overcome this apparent difficulty, one can perform an extra trivial Bogoliubov transformation of the quasi-particle operators of, e.g.\@ $| \Phi_k \rangle$, among themselves
\begin{align}
\begin{pmatrix}
\tilde{\beta} \\
\tilde{\beta}^{\dagger}
\end{pmatrix}_{[k]}
&\equiv  {\cal K}^{\dagger}_{[k]}
\begin{pmatrix}
\beta \\
\beta^{\dagger}
\end{pmatrix}_{[k]}   
= 
\begin{pmatrix}
K^{\dagger} & 0 \\
0 &  K^{T}
\end{pmatrix}_{[k]}
\begin{pmatrix}
\beta \\
\beta^{\dagger}
\end{pmatrix}_{[k]}
 \, , \label{furthertransfo1B}
\end{align}
where $K$ is a $n \times n$ unitary matrix. Maintaining the constraint from the phase convention, such a trivial Bogoliubov transformation modifies non-trivially the operator $S[k,l]$ and the auxiliary manifold linking both states without changing their overlap~\cite{Bally:2017nom}. Figure~\ref{results10levelsBCSmodel} illustrates this powerful flexibility of the method that can be used to bypass rare problems associated with potential zeros of the overlap along the path linking both states.

{\it Conclusions}. The present paper proposes a powerful method to compute the overlap between many-body states belonging to a set $\mathcal{M}$ used to represent the A-body Schr\"{o}dinger equation. Solving the corresponding secular equation requires the evaluation of the norm matrix and thus the capacity to compute its entries consistently and without any phase ambiguity. This is not always a trivial task, e.g.\@ it remained a long-standing problem for methods based on general Bogoliubov product states in used in nuclear physics. While a solution to this problem was found recently~\cite{Robledo:2009yd,Robledo:2011ce,Avez:2011wr}, the presently proposed method provides an alternative that can be generically applied to other classes of states of interest in many-body physics.

The overlap of any two states belonging to $\mathcal{M}$ is given as the exponential of the integral of the off-diagonal linked-connected kernel of an operator along an auxiliary continuous set joining both states. Such a linked-connected kernel is free from any phase ambiguity. The operator in question, which needs to be known or extracted during the procedure, is the hermitian generator of a unitary transformation connecting both states. 

In the present paper, the algebra is specified for sets made out of Bogoliubov states and numerically illustrated on the basis of a toy model. In this context, the overlap between arbitrary Bogoliubov states is computed, without any phase ambiguity, via elementary linear algebra operations. The numerical application nicely demonstrates the versatility of the method.

\begin{acknowledgments}
The authors thank J.-P. Ebran, M. Bender and M. Drissi for fruitful discussions as well as P. Arthuis, J. Bonnard, M. Drissi and V. Som\`a for proofreading the manuscript.
\end{acknowledgments}

\bibliography{overlap_PRL}

\end{document}